\documentclass[conference]{IEEEtran}
\IEEEoverridecommandlockouts
\usepackage{cite}
\usepackage{amsmath,amssymb,amsfonts}
\usepackage{algorithmic}
\usepackage{algorithm}
\usepackage{graphicx}
\usepackage{textcomp}
\usepackage{xcolor}
\usepackage{subfigure}
\usepackage{comment}
\usepackage{flushend}
\def\BibTeX{{\rm B\kern-.05em{\sc i\kern-.025em b}\kern-.08em
    T\kern-.1667em\lower.7ex\hbox{E}\kern-.125emX}}
\begin{document}

\title{Reconfigurable Intelligent Surface Assisted Semantic Communication Systems
}

	\author{
		\IEEEauthorblockN{Jiajia Shi\IEEEauthorrefmark{1}, Tse-Tin Chan\IEEEauthorrefmark{2}, Haoyuan Pan\IEEEauthorrefmark{3}, Tat-Ming Lok\IEEEauthorrefmark{1}}
		\IEEEauthorblockA{\IEEEauthorrefmark{1} Department of Information Engineering, The Chinese University of Hong Kong, Hong Kong SAR, China}
		\IEEEauthorblockA{\IEEEauthorrefmark{2} Department of Mathematics and Information Technology, The Education University of Hong Kong, Hong Kong SAR, China}
		\IEEEauthorblockA{\IEEEauthorrefmark{3} College of Computer Science and Software Engineering, Shenzhen University, Shenzhen, China}    
		\IEEEauthorblockA{E-mails: sj022@ie.cuhk.edu.hk, tsetinchan@eduhk.hk, hypan@szu.edu.cn, tmlok@ie.cuhk.edu.hk}
	}

\maketitle

\begin{abstract}

Semantic communication, which focuses on conveying the meaning of information rather than exact bit reconstruction, has gained considerable attention in recent years. Meanwhile, reconfigurable intelligent surface (RIS) is a promising technology that can achieve high spectral and energy efficiency by dynamically reflecting incident signals through programmable passive components. In this paper, we put forth a semantic communication scheme aided by RIS. Using text transmission as an example, experimental results demonstrate that the RIS-assisted semantic communication system outperforms the point-to-point semantic communication system in terms of bilingual evaluation understudy (BLEU) scores in Rayleigh fading channels, especially at low signal-to-noise ratio (SNR) regimes. In addition, the RIS-assisted semantic communication system exhibits superior robustness against channel estimation errors compared to its point-to-point counterpart. RIS can improve performance as it provides extra line-of-sight (LoS) paths and enhances signal propagation conditions compared to point-to-point systems.

\end{abstract}

\section{Introduction}\label{INTRO}

In recent years, semantic communication has gained significant attention as a solution to the semantic and effectiveness problems in post-Shannon communication system design \cite{Gündüz2023}. The concept of semantic communication can be traced back to Weaver's seminal work in \cite{weaver1949}. Building upon Weaver's ideas, Carnap and Bar-Hillel introduced semantic information theory in 1952, characterizing semantic information in a message \cite{carnap1952}. Unlike traditional communication systems that aim to recover exact bits or symbols, semantic communication aims to transmit only the relevant information or semantics of the transmitted signals, thus reducing data traffic and improving transmission efficiency.

The rapid development of deep learning (DL) has sparked interest in DL-based semantic communication as a promising communication paradigm.  Most prior works have been carried out on end-to-end communication systems, with different types of training data (also known as background knowledge) for different application scenarios. For example, DeepSC \cite{xie2021} is an essential work on semantic communication for text understanding based on Transformer. For image transmission that is semantically richer and more sensitive to bandwidth, a reinforcement learning-based semantic communication system was developed in \cite{huang2023}. Besides, for speech signals, an attention mechanism-based semantic communication system, named DeepSC-S, was presented in \cite{weng2021}. Compared to traditional methods, semantic communication systems are shown to be more robust to channel distortions, especially in low signal-to-noise ratio (SNR) regimes \cite{xie2021, weng2021}. For more examples of semantic communication systems, we refer the reader to \cite{Gündüz2023}.

Reconfigurable intelligent surface (RIS) has also gained significant attention due to their ability to manipulate the wireless medium by reflecting individual propagation paths through software control. An RIS is made up of numerous passive reflecting elements that can adjust the phase and amplitude of the incident signal \cite{liu2021}. Through collaboration among these elements, various paths can be tuned to enhance channel quality in various ways \cite{liaskos2018} without requiring additional power supply or complicated signal processing \cite{han2019, zhi2021}, making RIS highly desirable for practical implementation. 

Existing research on RIS focuses extensively on classical communication problems such as channel estimation and beamforming. For example, in \cite{wymeersc2021}, the authors proposed a two-stage channel estimation scheme that uses atomic norm minimization to sequentially estimate channel parameters. In \cite{demir2022}, several channel estimation schemes were derived for two RIS-assisted massive multiple-input multiple-output (MIMO) configurations. Furthermore, a novel beam training method with lower time overhead was introduced in \cite{hu2020}. Additionally, the authors in \cite{an2022} developed a low-complexity channel estimation and passive beamforming framework for RIS-based MIMO systems with discrete phase shifts at each reflecting element.

To the best of our knowledge, there is a lack of semantic communication models aided by RIS that can take unique properties of RIS to enhance channel quality. Inspired by the significant benefits of RIS, we develop an RIS-assisted semantic communication system. The main contributions of this work can be summarized as follows.

\begin{enumerate}
    \item We propose an RIS-assisted semantic communication system that leverages the advantages of RIS to improve the transmission performance of the semantic meaning over point-to-point semantic communication systems. By using DL, the transmitter, RIS, and receiver can be jointly optimized to cope with channel noise and semantic distortion.
    \item We evaluate the proposed system under Rayleigh fading channels with perfect and imperfect channel state information (CSI) and compare it with the point-to-point benchmark. Numerical results demonstrate that the RIS system achieves better bilingual evaluation understudy (BLEU) scores and is much closer to the transmission upper bound obtained by conveying noiseless text features to the receiver. Furthermore, it is shown that the RIS-assisted system is more robust to channel estimation errors in the case of imperfect CSI estimation compared to the point-to-point benchmark system.
\end{enumerate}

\textit{Notation}: $\mathbb{C}^{m\times n}$ represents sets of complex-valued matrices of size $m\times n$, and $j\triangleq\sqrt{-1}$ is the imaginary unit. $\mathbf{Y}\in\mathbb{R}^{a\times b}$ indicates that $\mathbf{Y}$ is an $a\times b$ matrix with real-valued elements. Given a vector $\mathbf{x}$, $x_i$ indicates its $i$-th component, and $|\cdot|$ denotes the modulus of a complex number. $\mathbb{E}[\cdot]$ denotes the statistical expectation, and $(\cdot)^T$ means the transpose. Single boldface letters represent vectors or matrices, and single plain letters denote scalars. $\mathbf{n}\sim\mathcal{CN}(0, \sigma^2)$ denotes zero mean circularly symmetric complex Gaussian noise vector $\mathbf{n}$ with variance $\sigma^2$. In addition, $\mathbf{A}={\rm diag}(\cdot)$ represents that $\mathbf{A}$ is a diagonal matrix with specified elements.

\section{System Model and Problem Formulation}\label{SMPF}

\begin{figure}[!t] 
\centerline{\includegraphics[width=0.5\textwidth]{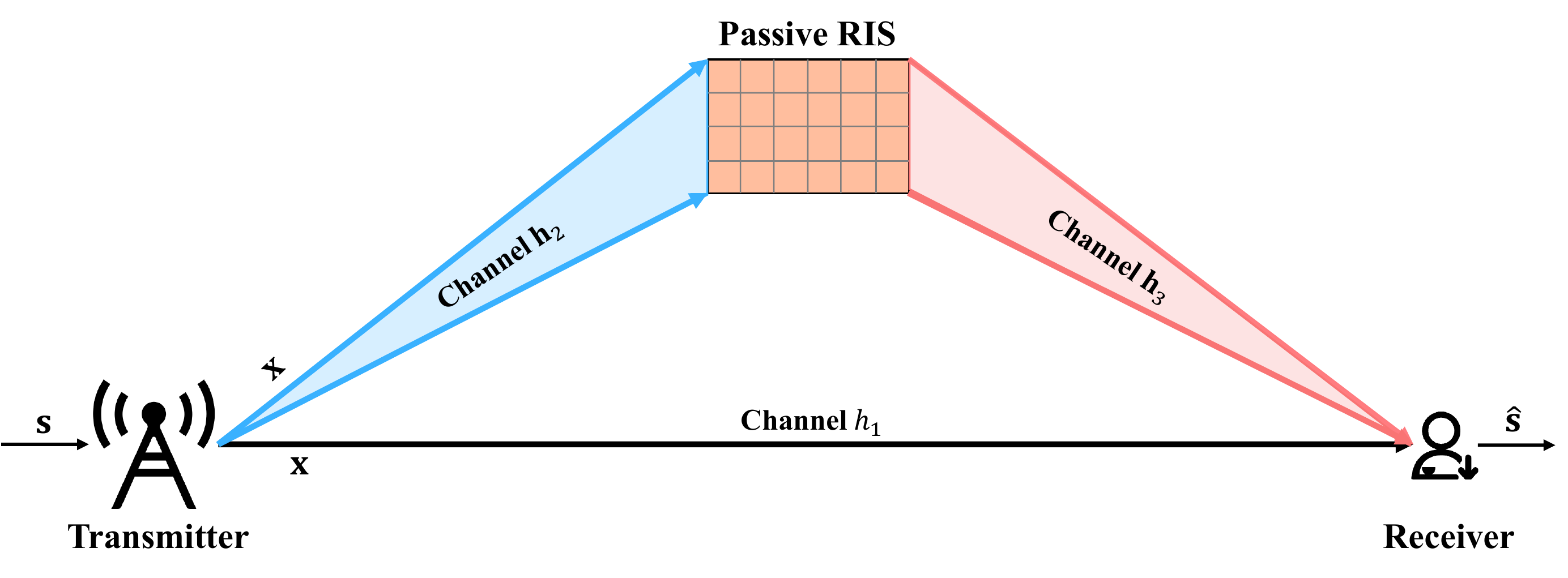}}
\caption{RIS-assisted semantic communication model.}
\label{fig1}
\end{figure}

As shown in Fig.~\ref{fig1}, we consider an RIS-assisted semantic communication system that consists of a transmitter, an RIS with $N$ individually passive reflecting elements, and a receiver. Both the transmitter and the receiver are equipped with a single antenna. Each reflecting element is capable of reflecting the incident signal with a reconfigurable phase shift and amplitude through an intelligent controller.

\subsection{Transmitter}\label{TX}

\begin{figure}[!t] 
\centerline{\includegraphics[width=0.5\textwidth]{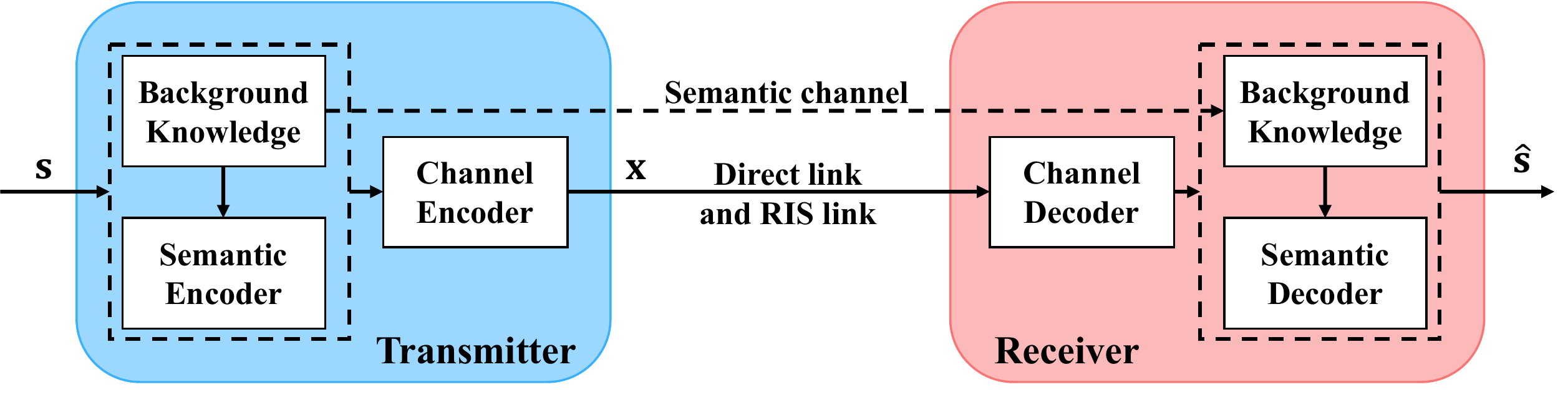}}
\caption{Basic framework of the RIS-assisted semantic communication system.}
\label{fig2}
\end{figure}

The transmitter takes a sentence, $\mathbf{s}=[w_1,w_2,...,w_L]$, as input, where $w_l$ represents the $l$-th word in the sentence. As illustrated in Fig.~\ref{fig2}, the transmitter consists of a semantic encoder and a channel encoder, which extract the semantics from $\mathbf{s}$ and ensure successful transmission of the semantics over wireless channels. The encoded symbol stream of the transmitter can be represented by
\begin{equation}
\mathbf{x}=\boldsymbol{E}^\mathcal{C}_{\boldsymbol{\beta}}(\boldsymbol{E}_{\boldsymbol{\alpha}}^\mathcal{S}(\mathbf{s})),
\label{eq1}
\end{equation}

\noindent where $\mathbf{x}\in\mathbb{C}^{1\times M}$ is normalized to the unit power, i.e., $\mathbb{E}[\mathbf{xx}^T]\leq 1$. The channel encoder with parameter set $\boldsymbol{\beta}$ is denoted by $\boldsymbol{E}^\mathcal{C}_{\boldsymbol{\beta}}(\cdot)$, while the semantic encoder with parameter set $\boldsymbol{\alpha}$ is represented by $\boldsymbol{E}^\mathcal{S}_{\boldsymbol{\alpha}}(\cdot)$.

\subsection{Reconfigurable Intelligent Surface (RIS)}\label{RIS}

To enhance communication between the transmitter and the receiver, a single RIS with $N$ passive reflective elements is employed in the proposed system, as depicted in Fig.~\ref{fig1}. Each reflective element comprises an atom that can adjust the phase and amplitude of each incident wave. The set of indices between 1 and $N$ is denoted by $\mathcal{N}$, i.e., $\mathcal{N} \triangleq \{1, 2, ..., N\}$. The reflection coefficient matrix of the RIS is represented by
\begin{equation}
    \mathbf{\Phi}\triangleq {\rm diag}(\gamma_1e^{j\phi_1},\gamma_2e^{j\phi_2},...,\gamma_Ne^{j\phi_N})\in\mathbb{C}^{N\times N},
    \label{eq2}
\end{equation}
where for all $n \in \mathcal{N}$, $\gamma_n \in [0, 1]$ and $\phi_n \in [0, 2\pi)$ denote the amplitude reflection coefficient and phase shift of the $n$-th element of the RIS, respectively.

\begin{figure*}[!t]
\centerline{\includegraphics[width=1\textwidth]{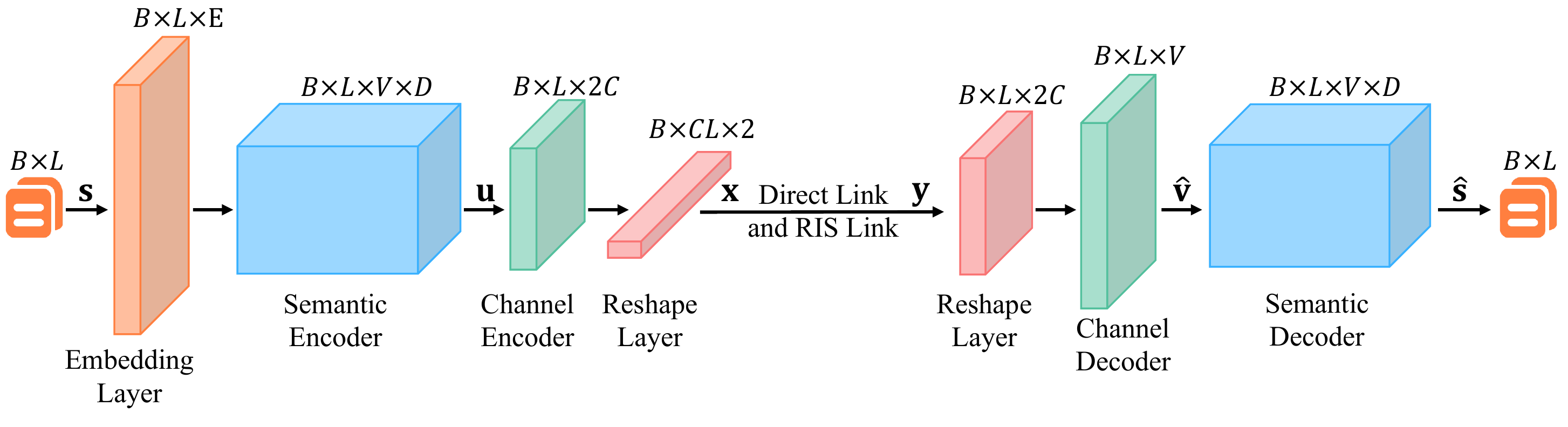}}
\caption{Neural network structure for the RIS-assisted semantic communication system.}
\label{fig3}
\end{figure*}

\subsection{Receiver}\label{RX}

The fading channel from the transmitter to the receiver is denoted by 
\begin{equation}
    h_1=|h_1|e^{j\theta_1},
    \label{eq3}
\end{equation}
where $|h_1|$ and $\theta_1$ are the channel amplitude and phase of $h_1$, respectively. The fading channel vectors of the transmitter-RIS and the RIS-receiver links are represented as
\begin{align}
    \mathbf{h}_2 &=[h_{21},h_{22},...,h_{2N}]^T\in\mathbb{C}^{N\times1},\ \text{and}
    \label{eq4} \\
    \mathbf{h}_3 &=[h_{31},h_{32},...,h_{3N}]^T\in\mathbb{C}^{N\times1},
    \label{eq5}
\end{align}
respectively. Here, $h_{2n}=|h_{2n}|e^{j\theta_{2n}}$ and $h_{3n}=|h_{3n}|e^{j\theta_{3n}}$ denote the $n$-th transmitter-RIS channel condition and the $n$-th RIS-receiver channel condition associated with the $n$-th reflecting element, respectively. The channel amplitude and phase of $h_{2n}$ are denoted by $|h_{2n}|$ and $\theta_{2n}$, respectively, while $|h_{3n}|$ and $\theta_{3n}$ represent the channel amplitude and phase of $h_{3n}$, respectively. It is assumed that $h_1$, $\mathbf{h}_2$, and $\mathbf{h}_3$ are independent and identically distributed (i.i.d.) fading channels. 

The information received at the receiver from the direct and the RIS paths can be expressed as
\begin{align}
\mathbf{y}&=(\underbrace{h_1+\mathbf{h}_3^T{\mathbf{\Phi}}\mathbf{h}_2}_{\Delta})\mathbf{x}+\mathbf{n},
\label{eq6}
\end{align}
where $\mathbf{n}\sim \mathcal{CN}(0,\sigma^2)$ is additive complex Gaussian noise with zero mean and variance $\sigma^2$.

The receiver consists of a channel decoder and a semantic decoder, as shown in Fig.~\ref{fig2}. The channel decoder alleviates channel attenuation and distortion, and the semantic decoder recovers the semantics of the sentences. The decoded information from the semantic decoder can be represented as
\begin{equation}
\widehat{\mathbf{s}}=\boldsymbol{D}^\mathcal{S}_{\boldsymbol{\chi}}(\boldsymbol{D}^\mathcal{C}_{\boldsymbol{\delta}}(\mathbf{y})),
\label{eq7}
\end{equation}
where $\boldsymbol{D}^\mathcal{S}_{\boldsymbol{\chi}}$ and $\boldsymbol{D}^\mathcal{C}_{\boldsymbol{\delta}}$ are the semantic decoder and the channel decoder with respect to parameters $\boldsymbol{\chi}$ and $\boldsymbol{\delta}$ in the RIS model, respectively. 

\section{Proposed Algorithm for RIS-assisted Semantic Communication System}\label{PA}

In this section, we first discuss the amplitude and phase shift of each reflection element, and then give the training process of the RIS system. After that, we present the performance metric used to evaluate the system.

\subsection{RIS Amplitudes and Phase Shifts Selection Scheme}\label{RISA}

The RIS reflects the signals to the receiver through the reflection coefficient matrix $\mathbf{\Phi}$ in \eqref{eq2}. To achieve this, we need to determine the amplitude reflection coefficient and phase shift of each element in the RIS. For fair comparisons, $\gamma_n$ is set to $1/N$. Our objective is to maximize the norm of $\Delta$ in \eqref{eq6}. Thus, we can obtain the each phase shift $\phi_n^*$ by solving problem $\mathcal{P}$,
\begin{equation}
\mathcal{P}: \mathop{\arg\max}_{\phi_n\in[0, 2\pi)} |h_1+\mathbf{h}_3^T\mathbf{Q}\mathbf{h}_2|.
\label{eq11}
\end{equation}

Solving problem $\mathcal{P}$ requires aligning the phases, and the required phase shift $\phi_n^*$ of the $n$-th element is given by
\begin{equation}
\phi_n^*=\theta_1-\theta_{2n}-\theta_{3n}.
\label{eq12}
\end{equation}
The required reflection coefficient matrix used in \eqref{eq2} and \eqref{eq6} can be expressed as
\begin{align}
\mathbf{\Phi^*}&={\rm diag}(\dfrac{1}{N}e^{j(\theta_1-\theta_{21}-\theta_{31})},\dfrac{1}{N}e^{j(\theta_1-\theta_{22}-\theta_{32})},\nonumber \\& \ \ \ \ ...,\dfrac{1}{N}e^{j(\theta_1-\theta_{2N}-\theta_{3N})}).
\label{eq13}
\end{align}
Determining the phase shifts requires the CSI of transmitter-RIS, transmitter-receiver, and RIS-receiver links. Besides, the received signal $\mathbf{y}$ in \eqref{eq6} needs to be multiplied by $e^{-j\theta_1}$ before the channel decoding is performed.

\subsection{Model Training}\label{MT}

\begin{algorithm}[!t]
\caption{Training Algorithm of the RIS-assisted Semantic Communication System.} \label{al1}
    \begin{algorithmic}[1]
        \STATE \textbf{Initialization}: Initialize the parameter sets $\boldsymbol{\beta, \alpha, \chi, \delta}$.
	\STATE \textbf{Input}: Knowledge set $\mathcal{K}$, training sentences $\mathbf{s}$ from training dataset, fading channel coefficients $h_1,\mathbf{h}_2$, and $\mathbf{h}_3$, and Gaussian noise $\mathbf{n}$.
        \WHILE {Stop criterion is not met}
        \STATE \textbf{Transmitter}:
            \STATE \ \ $\boldsymbol{E}_{\boldsymbol{\alpha}}^\mathcal{S}(\mathbf{s})\rightarrow \mathbf{u}$.
            \STATE \ \ $\boldsymbol{E}_{\boldsymbol{\beta}}^\mathcal{C}(\mathbf{u})\rightarrow \mathbf{x}$.
            \STATE \ \ Transmit $\mathbf{x}$ over the direct link and the transmitter-RIS\\ \ \ path.
        \STATE \textbf{RIS}: 
            \STATE \ \ Reflect the incoming signal according to the reflection \\ \ \ coefficient matrix in \eqref{eq13}.
        \STATE \textbf{Receiver}: 
            \STATE \ \ Receive $\mathbf{y}$ and multiply by $e^{-j\theta_1}$.
            \STATE \ \ $\boldsymbol{D}_{\boldsymbol{\delta}}^\mathcal{C}(\mathbf{y})\rightarrow \widehat{\mathbf{v}}$.
            \STATE \ \ $\boldsymbol{D}_{\boldsymbol{\chi}}^\mathcal{S}(\widehat{\mathbf{v}})\rightarrow \widehat{\mathbf{s}}$.
            \STATE \ \ Compute loss $\mathcal{L}_{CE}(\mathbf{s},\widehat{\mathbf{s}};\boldsymbol{\beta,\alpha,\chi,\delta})$ by \eqref{eq8}.
            \STATE \ \ Train $\boldsymbol{\beta,\alpha,\chi,\delta}$ with stochastic gradient descent.
        \ENDWHILE
        \STATE \textbf{Output}: The trained neural networks $\boldsymbol{E}_{\boldsymbol{\alpha}}^\mathcal{S}(\cdot),\boldsymbol{E}_{\boldsymbol{\beta}}^\mathcal{C}(\cdot),\boldsymbol{D}_{\boldsymbol{\delta}}^\mathcal{C}(\cdot)$, and $\boldsymbol{D}_{\boldsymbol{\chi}}^\mathcal{S}(\cdot)$.
    \end{algorithmic}
\end{algorithm}

The neural network structure for the RIS-assisted semantic communication system is shown in Fig.~\ref{fig3}. The system comprises four neural networks: semantic encoder, channel encoder, channel decoder, and semantic decoder. The semantic encoder and decoder are implemented using Transformer layers \cite{vaswani2017}, while the channel encoder and decoder are implemented using fully connected layers. 

The input to the RIS system is represented as $\mathbf{s}\in\mathbb{R}^{B\times L}$, where $B$ is the batch size and $\mathbf{s}$ is a set of text sample sequences. The input sequences are embedded into $\mathbb{R}^{B\times L\times E}$ for training, where $L$ is the number of embedded vectors, and $E$ is the length of each vector. Then, the semantic encoder learns features from embedding sequences to produce the output $\mathbf{u}\in\mathbb{R}^{B\times L\times V \times D}$. The channel encoder converts $\mathbf{u}$ to the space $\mathbb{R}^{B\times L\times 2C}$ and then reshapes it to the space $\mathbb{R}^{B\times CL\times 2}$ for transmission over the physical channel. The received signal, $\mathbf{y}$, is reshaped to the space $\mathbb{R}^{B\times L\times 2C}$ before being fed into the channel decoder. The output of the channel decoder is $\hat{\mathbf{v}}\in \mathbb{R}^{B\times L\times V}$, which is then converted to $\hat{\mathbf{s}}$ by the semantic decoder.

The goal of the system is to recover the semantics as accurately as possible, rather than successfully transmitting the original bits. To optimize at the semantic level, we use the cross-entropy loss as the cost function to minimize the differences between $\mathbf{s}$ and $\widehat{\mathbf{s}}$, which can be formulated as
\begin{align}
&\mathcal{L}_{CE}(\mathbf{s},\widehat{\mathbf{s}};\boldsymbol{\beta,\alpha,\chi,\delta})=\nonumber \\&-\sum_{l=1}q(w_l){\rm log}(p(w_l))+(1-q(w_l)){\rm log}(1-p(w_l)),
\label{eq8}
\end{align}
where $q(w_l)$ is the real probability of the $l$-th word $w_l$ appearing in sentence $\mathbf{s}$, and $p(w_l)$ denotes the predicted probability of the $l$-th word $w_l$ in sentence $\widehat{\mathbf{s}}$. Thus, the trainable parameters of the RIS system are updated simultaneously by calculating the loss at the receiver and backpropagating it to the transmitter.

The training algorithm of the RIS-assisted semantic communication system is detailed in Algorithm~\ref{al1}. The system initializes the parameters and uses embedding vectors to represent the input. Once the signals are encoded at the transmitter, they are simultaneously transmitted to both the RIS and the receiver. The RIS then reflects the incident signals to the receiver, which undertakes decoding processes on the overall received signals. We assume that the neural network components of the whole transceiver are differentiable with respect to the corresponding parameters, which can be jointly optimized using stochastic gradient descent to minimize the loss value in \eqref{eq8}. This enables the network to learn contextual semantics. After passing through the whole system, the input sentence, $\mathbf{s}$, is recovered into $\hat{\mathbf{s}}$. 

\subsection{Performance Metric}\label{PM}

In order to assess the accuracy of the RIS system's predictions, a quantitative metric is required. The bilingual evaluation understudy (BLEU) score is a widely used metric in machine translation research to measure the quality of results \cite{belghazi2018}.

In this paper, we use 1-gram and 2-gram BLEU scores to evaluate the performance of our model. Specifically, 1-gram means individual words in a sentence, while 2-gram refers to pairs of consecutive words. For instance, in the sentence ``I have an apple'', the 1-gram terms are ``I'', ``have'', ``an'', and ``apple'', while the 2-gram phrases are ``I have'', ``have an'', and ``an apple''.

For the target sentence $\mathbf{s}$ with length $l_{\mathbf{s}}$ and the predicted sentence $\boldsymbol{\widehat{\rm s}}$ with length $l_{\boldsymbol{\widehat{\rm s}}}$, the BLEU score can be computed by
\begin{equation}
{\rm log}\ {\rm BLEU}={\rm min}\left(1-\dfrac{l_{\boldsymbol{\widehat{\rm s}}}}{l_{\mathbf{s}}},0\right)+\sum_{i=1}^N w_i{\rm log}(p_i),
\label{eq9}
\end{equation}
where $w_i$ is the weight of the $i$-gram. Here $p_i$ is the $i$-gram score, which is defined as
\begin{equation}
p_i=\dfrac{\sum_k{\rm min}(C_k(\boldsymbol{\widehat{\rm s}}),C_k(\mathbf{s}))}{\sum_k{\rm min}(C_k(\boldsymbol{\widehat{\rm s}}))},
\label{eq10}
\end{equation}
where $C_k(\cdot)$ is the frequency count function for the $k$-th element in the $i$-gram. 

The BLEU score ranges from 0 to 1 and measures the similarity of the decoded text to the reference text. A score closer to 1 indicates a higher similarity, while 0 represents no similarity. In other words, in this paper, the higher the BLEU score obtained, the more accurately the semantics of the message is conveyed from the transmitter to the receiver.

\section{Experimental Results}\label{ENR}

In this section, we present experimental results that demonstrate the superiority of the RIS system. We first outline the experimental setup and then compare the performance in Rayleigh fading channels with perfect CSI. Afterward, we conduct evaluations in the presence of imperfect CSI estimation as a reference in practice.

\subsection{Dataset and Benchmarks}\label{DB}

\begin{figure}[!t]
	\centering
	\subfigure[]{
		\begin{minipage}[b]{0.5\textwidth}
			\includegraphics[width=1\textwidth]{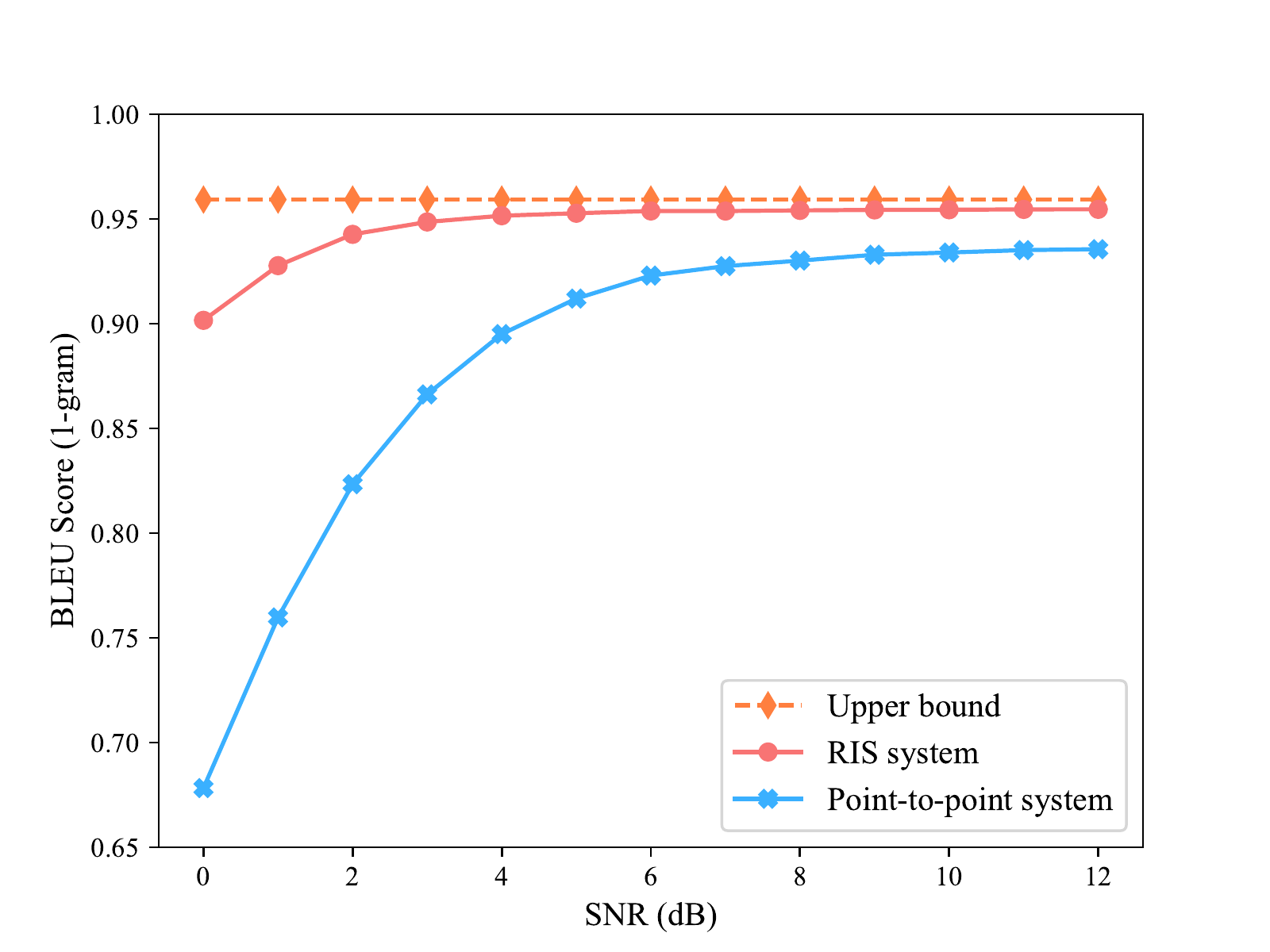}
		\end{minipage}
		\label{fig4a}
	}
        \\
    	\subfigure[]{
    		\begin{minipage}[b]{0.5\textwidth}
   		 	\includegraphics[width=1\textwidth]{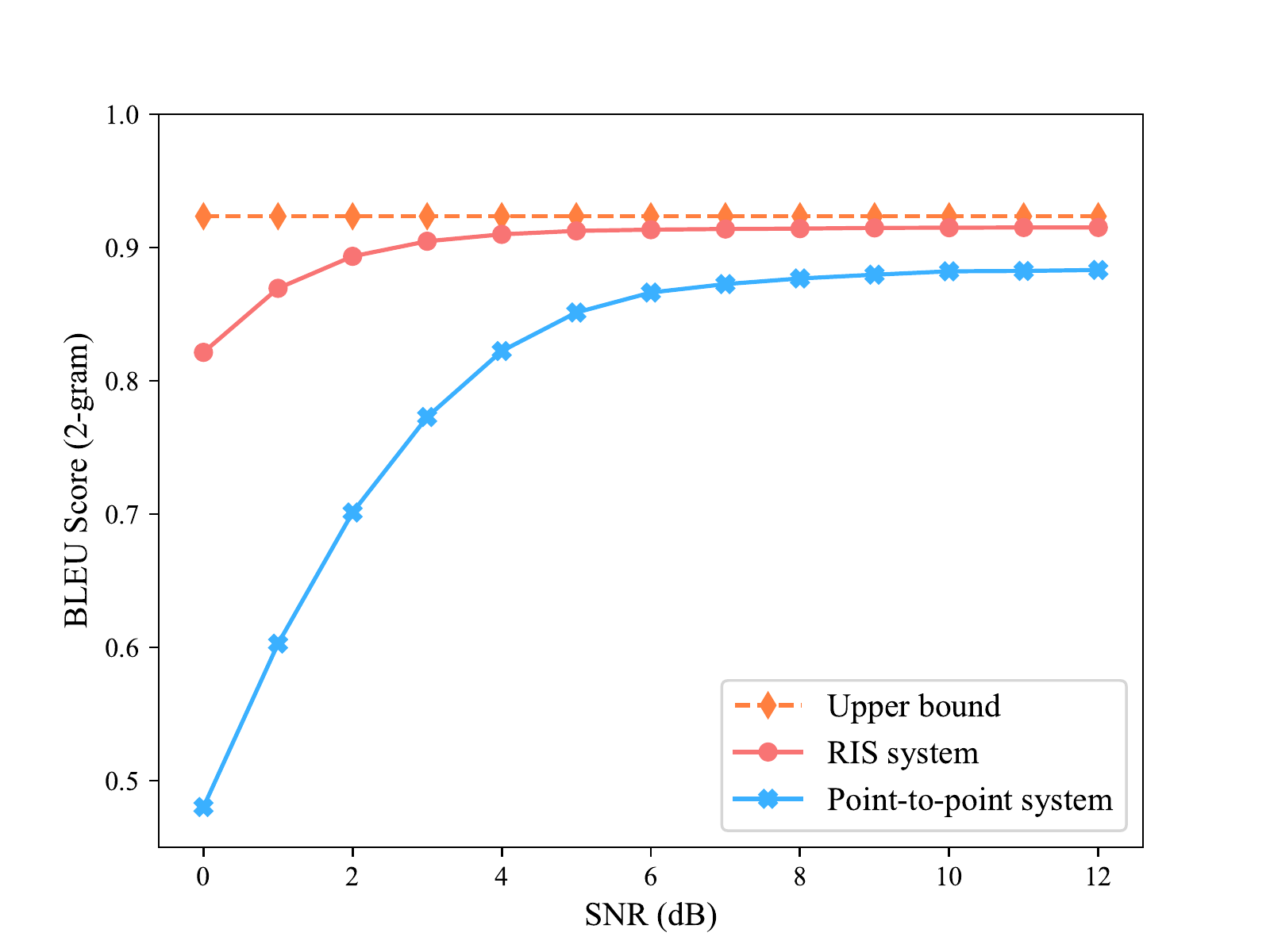}
    		\end{minipage}
		\label{fig4b}
    	}
	\caption{(a) 1-gram and (b) 2-gram BLEU scores versus SNR for our RIS system and two benchmarks over Rayleigh fading channels with perfect CSI estimation.}
	\label{fig4}
\end{figure}

The European Parliament proceedings dataset \cite{koehn2005}, which comprises approximately 2 million sentences and 53 million words, is used in our experiments. The investigated semantic communication systems are trained using a fixed signal-to-noise ratio (SNR) of $7$ dB and a random set of independent and identically distributed (i.i.d.) Rayleigh fading channels for each link. After training, the models are loaded to evaluate their performance under another set of Rayleigh fading channels and SNRs. In all experiments, we choose $N=10$ for the RIS system.

To provide performance comparisons, we use the following two benchmarks.
\begin{enumerate}
    \item \textit{\textbf{Point-to-point system}:} The first benchmark is to use the same setup without the RIS components, i.e., a point-to-point semantic communication system.
    \item \textit{\textbf{Upper bound}:} The second benchmark, referred to as the upper bound, utilizes the same transmitter and receiver as the point-to-point semantic communication system. However, this benchmark considers noiseless text features directly conveyed from the transmitter to the receiver.
\end{enumerate}

\subsection{Results Comparison}\label{RC}

\begin{figure}[!t]
	\centering
	\subfigure[]{
		\begin{minipage}[b]{0.5\textwidth}
			\includegraphics[width=1\textwidth]{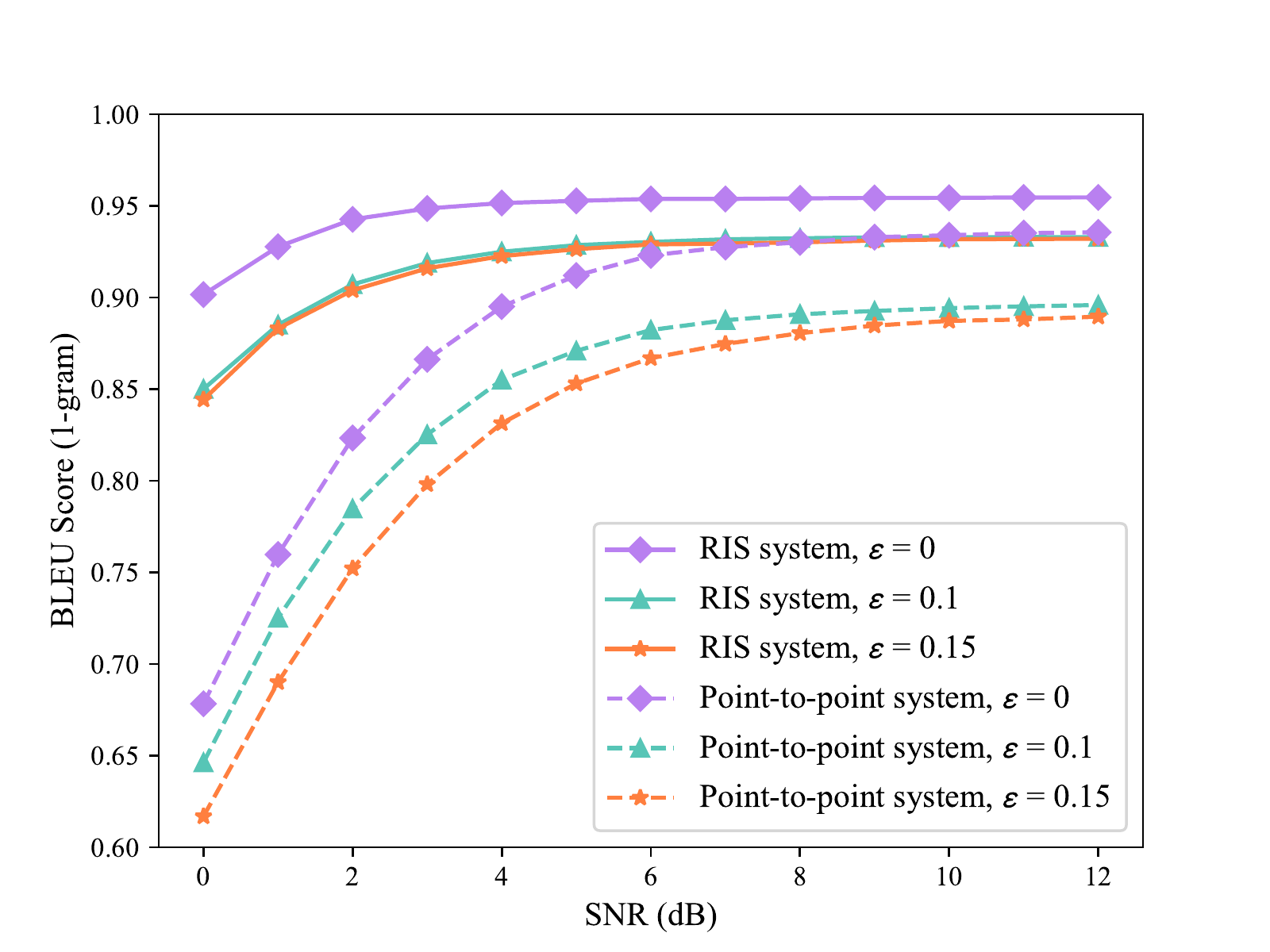}
		\end{minipage}
		\label{fig5a}
	}
        \\
    	\subfigure[]{
    		\begin{minipage}[b]{0.5\textwidth}
   		 	\includegraphics[width=1\textwidth]{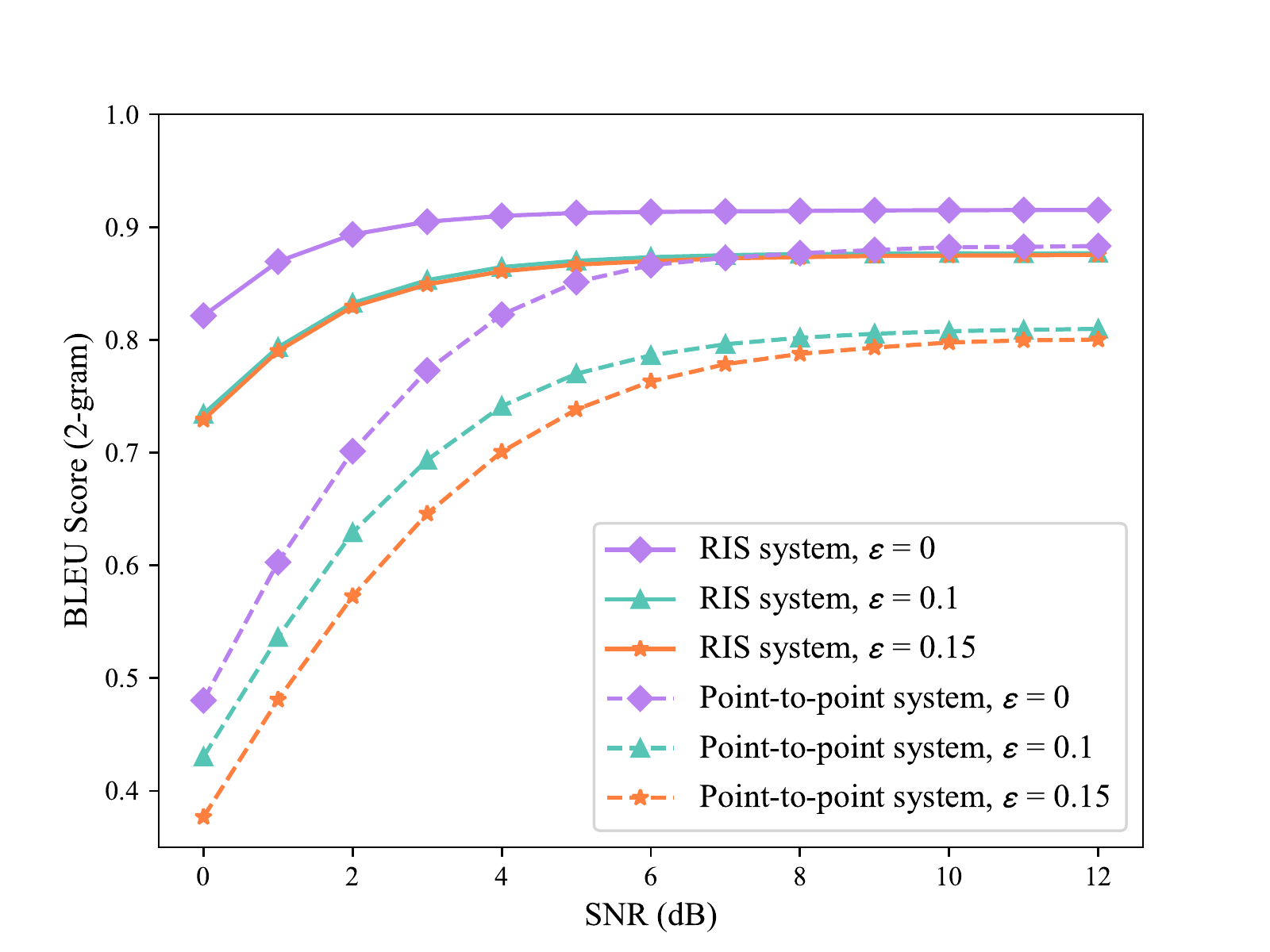}
    		\end{minipage}
		\label{fig5b}
    	}
	\caption{(a) 1-gram and (b) 2-gram BLEU scores versus SNR for our RIS and point-to-point systems over Rayleigh fading channels with imperfect CSI estimation.}
	\label{fig5}
\end{figure}

Fig.~\ref{fig4} compares the 1-gram and 2-gram BLEU scores of the RIS system and the two benchmarks under Rayleigh fading channels, assuming perfect CSI is known at the transmitter, RIS, and receiver. The results show that the RIS system outperforms the point-to-point system in the tested channel environments, particularly in the low SNR regime. Additionally, the BLEU scores of the RIS system are closer to the upper bound than those of the point-to-point system. Specifically, when the SNR is higher than $5$ dB under Rayleigh fading channels, the difference between the RIS system and the upper bound in the 1-gram (2-gram) BLEU score is less than $0.55\%$ ($1.01\%$). This indicates that the RIS system can recover the semantics of the transmitted text more effectively than the point-to-point semantic communication system, especially in the low SNR regime.

In practical scenarios, channel estimation is often imperfect. Hence, we also perform experimental evaluations with imperfect CSI estimation, where the estimated channel can be modeled as 
\begin{equation}
\mathbf{\hat{h}}=\mathbf{h}(1+\mathbf{e}),
\label{eq14}
\end{equation}
where $\mathbf{h}$ and $\mathbf{\hat{h}}$ are the actual and estimated channel coefficients, respectively, and $\mathbf{e}\sim\mathcal{CN}(0, \epsilon^2)$ is the channel estimation error. 

Fig. \ref{fig5} evaluates the BLEU scores of the RIS and point-to-point systems under Rayleigh fading channels with varying $\epsilon$. As shown in the results, the performance of both systems degrades as $\epsilon$ increases (i.e., as the channel estimation error increases). However, the RIS system exhibits smaller performance degradation compared to the point-to-point system, indicating its greater robustness to channel estimation errors. 

The performance improvement of the RIS-assisted semantic communication system comes from the ability of RIS to manipulate the wireless environment to improve signal transmission. The RIS can create additional paths for signals, thereby increasing diversity, which can potentially improve the reliability and quality of communication. In contrast, the point-to-point semantic communication system has only a direct path between the transmitter and receiver. If this single path encounters significant fading or channel estimation error, it can degrade the quality of the communication.

\section{Conclusion}\label{CON}

In this paper, we introduce an RIS-assisted semantic communication system for text transmission. Our proposed system utilizes an RIS to enhance the received signal quality and improve the overall performance over point-to-point semantic communication systems. Experimental results demonstrate that the RIS system achieves better BLEU score performance, particularly in low SNR regimes. Furthermore, the experimental results indicate that the RIS system can still perform well under imperfect channel estimation, which highlights its potential for real-world applications. Overall, the RIS system provides a promising solution for improving the efficiency and reliability of semantic communication.


\vspace{12pt}

\end{document}